\documentclass[ twocolumn,amssymb,amsmath]{revtex4}

\usepackage{graphicx}
\usepackage{dcolumn}
\usepackage{bm}

\def\be{\begin{equation}}
\def\ee{\end{equation}}

\def\bm{\begin{displaymath}}
\def\ebm{\end{displaymath}}
\def\npm{n_{\pm \epsilon}}
\def\npe{n_{\epsilon}}

\def\npepr{n_{\epsilon^\prime}}

\def\bq{{bf q}}
\newcommand{\frc}[2]{\frac{\partial #1}{\partial #2}}
\newcommand{\bea}{\begin{eqnarray}}
\newcommand{\eea}{\end{eqnarray}}

\newcommand{\dineq} [7] {\frac{\partial #1} {\partial #2} =
#3^{#4}\left( #1  \right)+#5^{#6} \left( #1,#7 \right) }
\newcommand{\dineqPh} [6] { \frac{\partial #1}{\partial#2} =
#3\left( #1  \right)+#4^{#5} \left( #1,#6 \right) }
\newcommand{\Icol} [3] { I^{#3}\left( #1, #2 \right) }
\newcommand{\Qcol} [2] { Q^{#2}\left( #1 \right) }
\newcommand{\Lcol} [1] { L\left( #1 \right) }

\begin{document}

\title{Phonon deficit effect and solid state refrigerators based on
superconducting tunnel junctions.}

\author{G.G.~Melkonyan$^{a}$, H.~Kr\"oger$^{a}$ and A.M.~Gulian$^{b}$}

\address{$^{a}$ D\'epartement de Physique Universit\'e Laval,
        Qu\'ebec, Qu\'ebec G1K 7P4, Canada \\
        $^{b}$ Naval Research Laboratory, Washington DC 20375, USA }

\begin{abstract}
Since the first demonstration of electron cooling in
super\-con\-duc\-tor--insulator--normal metal ($SIN$) tunnel junctions,
there is growing interest in tunneling effects for the purpose to develop
on--chip refrigerators which can generate cooling in the order of $100mK$ in
an environment of $0.3$--$0.5K$. Thin film devices have the advantage of
being extremely compact, operate in a continuous mode, dissipate little
power, and can easily be integrated in cryogenic detectors. Motivated by
such possibilities, we investigate the phonon deficit effect in thin film $%
SIS$ (super\-con\-duc\-tor--insulator--super\-con\-duc\-tor) and
$SIN$ tunnel junctions. Under certain circumstances, the phonon
absorption spectra of such tunnel junctions have spectral windows
of phonon absorption/emission. We propose to use phonon filters to
select the phonon absorbtion windows and thus to enhance the
cooling effect. Membranes attached to such tunnel junctions can be
cooled in this way more effectively. We discuss a particular
superlattice design of corresponding phonon filters.
\end{abstract}

\maketitle

\section{Introduction}

Solid state micro--refrigerators ($SSMR$) have many advantages
over conventional cooling devices: they have a long life-time,
they are compact and simple to apply. These $SSMRs$ utilize motion
of electrons rather than atoms and molecules. The flow of electric
current through a semiconductor--semiconductor contact is
accompanied by a transfer of heat (the Peltier effect) from one
semiconductor to the other. Micro-refrigerators based on the
Peltier effect are working at temperatures above $150K$.
Solid-state micro-refrigerators operating below $150K$ are a
scientific and technical challenge. A possible approach to this
problem is
to consider electron kinetics in superconductor--insulator--normal metal ($%
SIN$) or superconductor--insulator--superconductor ($SIS$) tunnel junctions~%
\cite{caspek,Martinis}. As we will demonstrate in this article,
another even deeper understanding of the same effects is possible
 from the concept of the phonon deficit effect ($PDE$). The phonon
deficit effect was initially
predicted~\cite{GulianZharkov2,GulianZharkov1} to occur when
superconductivity  is enhanced  in high frequency electromagnetic
fields ($UHF$)~\cite{eliash}. Later the $PDE$ was recognized to
exist in more general cases~\cite{GulBook,GulianZharkov3}. It is
related to the violation of detailed balance of kinetic processes
in superconductors in the presence of an external field. If a
superconducting film is immersed into a heatbath and the electron
system goes over to an excited nonequilibrium state due to an
external supply of energy, then under appropriate conditions the
superconducting film absorbs phonons from the heatbath rather than
emitting phonons.
\begin{figure}[tbh]
\par
\begin{center}
\leavevmode
\includegraphics[width=0.65\linewidth]{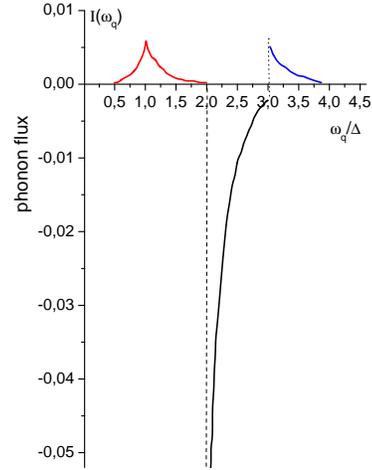}
\end{center}
\caption{Spectral dependence of the phonon flux (in units of energy per unit
time) radiated from a thin superconducting film (with smaller gap $\Delta
^{S}$) in an $SIS^{\prime }$ tunnel junction (taken from Ref.~\protect\cite
{GulBook}) }
\label{figsub_gul_absorp}
\end{figure}

\bigskip

The situation is essentially similar in the case of thin tunnel
junctions of superconductors. Phonon emission of a symmetric $SIS$
tunnel junction has been considered in detail in
Ref.~\cite{GulianZharkov3} and an asymmetric junction has been
considered in Refs.~\cite
{GulBook,GulianSergoyan,GulianMelkonyan}. In the sub--threshold
tunneling regime ($V<2\Delta /e$, where $\Delta $ denotes the gap)
the resulting phonon flux is shown in
Fig.[\ref{figsub_gul_absorp}]. One observes the phonon emission in
the frequency windows $0<\omega _{q}<2\Delta /\hbar $ and $\omega
_{q}>3\Delta /\hbar $. In the same time there is the phonon
absorption in the frequency window $2\Delta /\hbar<\omega _{q}<
3\Delta /\hbar$. The analogy between this case and the
superconductor in the $UHF$ field [see Ref.~\cite {GulBook},
Figure 6.2] strikingly demonstrates the universality of the
mechanism of the phonon deficit effect.

Experiments on refrigeration are usually performed on the $SIN$
tunnel junctions. So called ``cooling by extraction'' is
recognized as the operational principle of these type of
refrigerators. At temperatures much below the critical temperature
$T_{c}$ of the superconducting electrode ($S$) there are few
thermally excited quasiparticles above the superconducting gap
$\Delta $, so that tunneling from $S$ to the normal metal
electrode ($N$) can be neglected. Meanwhile, in the normal metal
electrode ($N$) there are enough electron-like and hole-like
thermal excitations near the Fermi energy $E_{F}$ which can tunnel
to $S$ and effectively modify the electron distribution in $N$.
The rate of electronic heat exchange with the lattice is
proportional to $T_{e}^{5}-T_{ph}^{5}$, where $T_{e}$ is
electronic temperature and $T_{ph}$ is the temperature of phonons.
The cooling process is a balance between phonon heat flow,
quasiparticle extraction and Joule heating. The value of the
applied potential $V$ on the tunnel junction determines if cooling
or heating occurs depending on which process is dominant in the
heat transfer balance.

Here we have studied the behavior of $SIN$ tunnel junction using
the phonon deficit effect. The outline of this paper is as
follows. In Section~\ref{heatbath} we describe the phonon heatbath
model and decoupling of electrons from phonons. In
Section~\ref{SIN_tunnel} we solve kinetic equations for the $SIN$
tunnel junction. We calculate the net phonon flux to the $SIN$
tunnel junction when electron and phonon systems are maintened at
different temperatures. We show that the rule for heat exchange
$P\propto T_{e}^{5}-T_{ph}^{5}$ follows from the $PDE$ in $SIN$
tunnel junctions. In analyzing the $PDE$ we propose the phonon
filters. We present the results of our numerical simulations,
which can be utilised to enhance the refrigeration effect. The
constraction of phonon filters in accordance with the calculated
phonon absorbtion spectra is discussed in
Section~\ref{PhononDeficitMicroRef}. Conclusions are given in
Section~\ref{Concl}.

\section{Phonon heatbath model and decoupling of electrons from phonons \label{heatbath} }

Experimentally junction electrodes are thin films placed on a substrate.
These thin films should be considered as a system of two interacting
subsystems: electrons and phonons. Nonequilibrium dynamics of phonons and
electrons in superconductors is described by a coupled system of kinetic
equations~\cite{GulBook}. In the simplest case, when the lattice phonons
play the role of a thermostat, the kinetic equations for electron (hole)
excitations and phonons in a metal can be written as in the following form~%
\cite{GulBook} \bea \label{ElectronKineticHeat} \dineq{n_{\pm \epsilon }} {t}
{Q} {f} {I} {el-ph} {N_{\omega _{q}}^{i}}, \\
\label{PhononKineticHeat} \dineqPh {N_{\omega _{q}}^{i}} {t} {L} {I} {ph-el}
{n_{\pm \epsilon }} ~ . \eea Here $n_{\pm \epsilon }$ denotes the
distribution functions of nonequilibrium electron-like ($n_{\epsilon }$),
respectively hole-like ($n_{-\epsilon }$) excitations in the film. $%
N_{\omega _{g}}^{i}$ represents the distribution function of internal
phonons, $I^{el-ph}(n_{\pm \epsilon },N_{\omega _{q}}^{i})$ stands for the
electron--phonon collision integral, and $I^{ph-el}(N_{\omega
_{q}}^{i},n_{\pm \epsilon })$ denotes the phonon--electron collision
integral. $\Qcol{n_{\pm \epsilon }}{f}$ is an external source of
nonequilibrium quasiparticles and $f$ represents external fields to be
specified later. $L(N_{\omega _{q}}^{i})$ is an operator describing the
interaction of phonons with an external heatbath.

\bigskip

An important problem in a dynamical theory of metals is the study
of nonequilibrium stationary states, i.e. when $\partial n_{\pm
\epsilon }/\partial t=0$, $\partial N_{\omega _{q}}^{i}/\partial
t=0$. For thin films in an external field the dynamical problem
can be simplified by considering the phonon heatbath
mo\-del~\cite{GulBook,Eliashberg} and also the
''geometric-acoustical'' approximation of phonon propagation. In
this
approximation the distribution function of internal phonons is given by $%
N_{\omega _{q}}^{i}=N_{\omega _{q}}^{0}$, where $N_{\omega _{q}}^{0}$ is the
distribution function of phonons at thermodynamic equilibrium. The
distribution of electrons and holes corresponds to a nonequilibrium state.
In a steady state, using the phonon heatbath model, the following system of
equations holds
\bea
\label{ElectronKineticHeat1}
-\Qcol{n_{\pm \epsilon}}{f}&=\Icol{n_{\pm\epsilon}}{N^0_{\omega_q}}{el-ph}, \\
 \label{PhononKineticHeat1}
-\Lcol{N^0_{\omega_q}}&=\Icol{N^0_{\omega_q}} {n_{\pm \epsilon}}{ph-el} ~ .
\eea
These equations are uncoupled. Eq.(\ref{ElectronKineticHeat1}) can
be used to obtain stationary steady state solutions for the
distribution functions of nonequilibrium excitations (electron and
hole branches. Eq.(\ref{PhononKineticHeat1}) becomes an identity,
which describes the phonon exchange (phonon flux) between the thin
film and its environment~\cite{GulBook}.

\section{$SIN$ tunnel junction - kinetic equations in the steady state \label%
{SIN_tunnel}}

We consider now a massive superconductor (injector $S^{\prime }$), having a
superconducting gap $\Delta ^{\prime }$, joined with a thermostat held at
temperature $T$. A thin film having a thickness $d\sim \xi _{0}$
(superconductor or normal metal) is attached to the massive superconductor
by means of a thin oxide layer. It has the same temperature (see Fig.[\ref
{fig_refr_design}]). In the dissipative steady state, nonequilibrium
electron ($n_{\epsilon }$) and hole ($n_{-\epsilon }$) excitations in the
thin film ($S$ or $N$) are described by the following kinetic equation (see
Eqs.~(\ref{ElectronKineticHeat}) and~(\ref{ElectronKineticHeat1}) )
\begin{equation}
u_{\epsilon }\frc{n_{\pm\epsilon}}{t}=Q(n_{\pm \epsilon })+I^{el-ph}(n_{\pm
\epsilon },N_{\omega }^{0})=0~.  \label{particle}
\end{equation}
Here $u_{\epsilon }=|\epsilon |\theta (\epsilon ^{2}-\Delta ^{2})/\sqrt{%
\epsilon ^{2}-\Delta ^{2}}$ denotes the BCS density of states, $\epsilon $
is the quasiparticle energy and $Q(n_{\pm \epsilon })\equiv \Qcol{n_{\pm
\epsilon }}{f}$ describes the injector. In a particular geometry the
external field $f$ represents the distribution function of quasiparticles in
the electrode denoted by $S^{\prime }$. The explicit form of the
electron--phonon collision integral $I^{el-ph}(\npe,N_{\omega _{q}}^{0})$
and $Q(n_{\pm \epsilon })$ are given in Refs.~\cite{GulBook,GulianZharkov1}.
The dissipative steady state solutions of Equation~(\ref{particle}) give the
electron and the hole distribution functions in the thin film electrode. The
corresponding phonon fluxes emitted to (or absorbed from) its environment in
a frequency interval $d\omega _{q}$ centered at frequency $\omega _{q}$ can
be obtained from (\ref{PhononKineticHeat1}).

\bigskip

The number of phonons absorbed per time unit in the volume $\Omega$ at
frequency $\omega_q$ in the spectral interval $\mathrm{d}\omega_q$ is given
by \be \label{PhononNumber} {\textrm{d}}\dot{N}_{\omega_q}=-%
\Lcol{N^0_{\omega_q}}\rho_0(\omega_q){\textrm{d}} \omega_q=
\rho_0(\omega_q)I^{ph-el}\left(N^0_{\omega_q}\right) {\textrm{d}} \omega_q ~
. \ee
Here $\rho_0(\omega_q)={\Omega} \omega_q ^2/(2\pi^2 u^3)$, $u$ is the
velocity of sound and $\dot{X} \equiv \partial X/\partial t$. The absorbed
power per volume $\Omega$ at frequency $\omega_q$ in the spectral interval $%
\mathrm{d}\omega_q$ is given by
\begin{equation}  \label{PhononEmmison}
\frac{{\mathrm{d}}P}{{\mathrm{d}}\omega_q}= \hbar \omega_q ~
\rho_0(\omega_q) ~ I^{ph-el}(N_{\omega_q}^0) ~ .
\end{equation}
Introducing the operator $I(N^0_{\omega_q})$ via
\[
{I} (N^0_{\omega_q})=\frac{16\pi \epsilon_F }{\lambda \omega_D }\,\omega_q^3%
\mathit{I}^{ph-el}\left( N^0_{\omega_q}\right) ,
\]
allows to express the absorbed power (\ref{PhononEmmison}) in dimensionless
form
\begin{eqnarray}  \label{Power2}
&&\frac{{\mathrm{d}} P}{{\mathrm{d}}x}=\frac{{\ \Omega} \lambda \omega_D }{%
16\pi \epsilon_F u^3}\frac{(k_B T_{c^\prime})^5}{\hbar^3}{I}(N^0_{x})=P_0 {\
I}(N^0_{x})  \nonumber \\
&& P_0=\frac{{\ \Omega} \lambda \omega_D }{16\pi \epsilon_F u^3}\frac{(k_B
T_{c^\prime})^5}{\hbar^3} ~ ,
\end{eqnarray}
where we abbreviate $x=\hbar\omega_q/(k_B T_{c^\prime})$ and $T_{c^\prime}$
denotes the critical temperature of the injector.

In the following we use the units $k_B=\hbar=e=1$ if not stated otherwise.
Also, recall that the primed variables correspond to the injector material.

\subsection{Numerical results and analysis}

To simplify further the analysis, we need to analyze properties of equation~(%
\ref{particle}). The electron--phonon collision integral
$I^{el-ph}$ is proportional to the parameter $\gamma \propto
T^3/\omega _D^2$ (damping) which characterizes the strength of the
relaxation processes. The parameter $\nu $ describes the intensity
of the particle and hole injection $Q(\npm)$. The typical values
of $\nu $ are small compared to $\gamma $. When $\nu /\gamma $
tends to zero, the particle and hole distribution functions go
over to
thermal equilibrium distribution functions and both, $I^{el-ph}(\npm%
,N_{\omega _q}^0)$ and $I^{ph-el}(N_{\omega _q}^0,\npm)$ vanish. At finite
values of $\nu /\gamma $ the particle and hole distribution functions differ
from the thermal equilibrium functions and thus phonon fluxes arise. As
observed in Refs.~\cite{GulianSergoyan,GulianZharkov1,GulianMelkonyan} both
negative and positive phonon fluxes occur between the thin film and the
thermal reservoir. The net phonon flux integrated over all frequencies can
be either negative or positive.

\bigskip

\begin{figure}[tph]
\par
\begin{center}
\leavevmode
\includegraphics[width=0.7\linewidth]{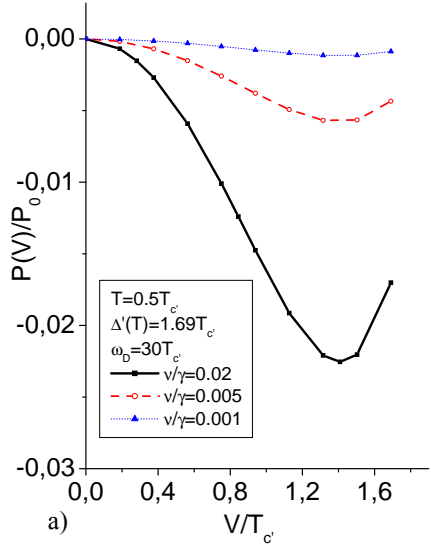} %
\includegraphics[width=0.7\linewidth,viewport= 13 13 599 820]{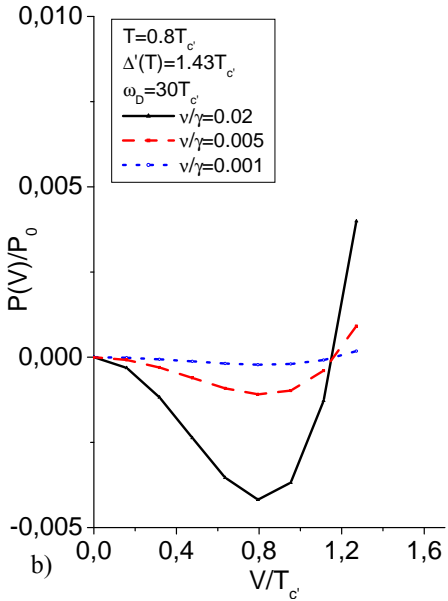}
\end{center}
\caption{Absorbed power $P(V)$, in units of $P_{0}$, by the normal metal
electrode of the thin film as function of applied potential for different
values of $\nu /\gamma $. $\Delta ^{\prime }(T)$ and $T_{c^{\prime }}$ are
the superconducting gap and the critical temperature of the injector,
respectively, $\omega _{D}$ is the Debye frequency of the normal metal
electrode of the thin film and $V$ is the applied potential across the
tunnel junction. (a) $T/T_{c^{\prime }}=0.5$, (b) $T/T_{c^{\prime }}=0.8$.
The lines are a guide to the eye. }
\label{AbsorbedT5}
\end{figure}

\begin{figure}[tbh]
\par
\begin{center}
\leavevmode
\includegraphics[width=0.7\linewidth]{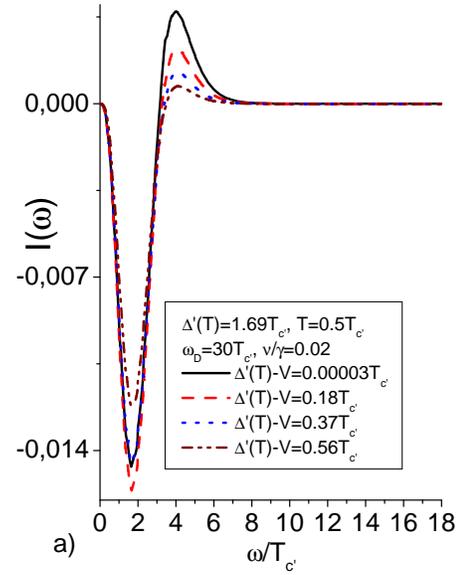} %
\includegraphics[width=0.7\linewidth, viewport= 13 13 599 820]{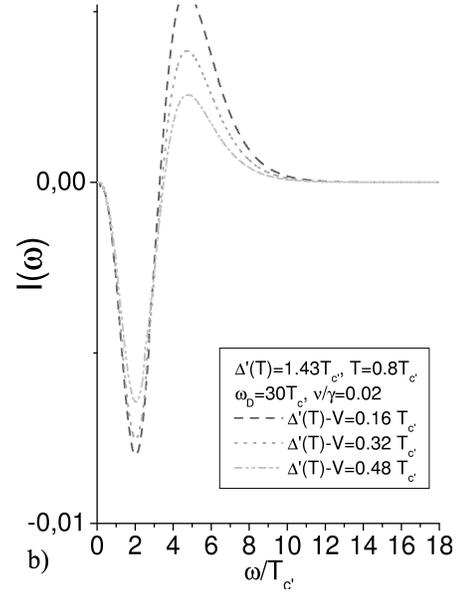}
\end{center}
\caption{Spectral absorption (negative part) and emission (positive part) of
phonons near the maximal phonon absorption potential of the thin film. $%
\Delta ^{\prime }(T)$ and $T_{c^{\prime }}$ are the superconducting gap and
the critical temperature of the injector, respectively. $\omega _{D}$ is the
Debye frequency of the normal metal electrode of the trhin film and $V$ is
the applied potential across the tunnel junction. (a) $T/T_{c^{\prime }}=0.5$%
, (b) $T/T_{c^{\prime }}=0.8$. }
\label{SpectralAbsorption}
\end{figure}

Figure [\ref{AbsorbedT5}] shows the power $P(V)/P(0)$ absorbed by the normal
metal thin film electrode. It is shown as a function of the potential
applied across the $S^{\prime }IN$ tunnel junction for different values of
the parameter $\nu /\gamma $. All curves have a minimum (maximal absorption)
at $V<\Delta ^{\prime }(T)$. When the potential difference $V$ across the
tunnel junction approaches the superconducting gap $\Delta ^{\prime }(T)$,
the phonon absorption decreases. The absorbed power increases with the
increasing values of $\nu /\gamma $. The situation changes dramatically in
the high temperature regime $T/T_{c^{\prime }}>0.5$. When the value of the
potential applied across the tunnel junction approaches the superconducting
gap $\Delta ^{\prime }$, then the $S^{\prime }IN$ tunnel junction starts to
emit phonons to the environment. This is shown in Fig.[\ref{AbsorbedT5}b].
In Fig.~[\ref{SpectralAbsorption}] we present the dynamics of phonon flux in
the vicinity of the minima of Figs~[\ref{AbsorbedT5}a,b]. As expected from
Fig.[\ref{AbsorbedT5}a], the phonon absorption spectrum is spread out over
the whole frequency range of acoustic phonons at the temperature $%
T/T_{c^{\prime }}=0.5$ and $\Delta ^{\prime }-V>0.18$. At temperature $%
T/T_{c}^{\prime }=0.8$ there is both phonon emission and phonon absorption
at high and low frequency regions, respectively.
The phonon emission spectrum of a $SIN$ tunnel junction is
different from the $S^{\prime }IS$ case (see
Fig.[\ref{figsub_gul_absorp}]). The $SIS$ tunnel junction has two
emission and one absorption frequency regions. In $SIN$ tunnel
junctions the low frequency emission region is absent. Also, as
temperature decreases the phonon emission decreases faster than
the absorption.

\subsection{$PDE$ in the $SIN$ tunnel junctions:''Equilibrium'' limit}

An analytic expression for the power absorbed by the $SIN$ tunnel
junction when electron and phonon subsystems are maintained at
temperature $T_{e}$ and $T_{ph}$, respectively, can be obtained by
integrating expression~(\ref {Power2}). The heatflow from the
phonon subsystem to the electron subsystem maintained at different
temperatures is given then by~(for details see Appendix ) \be
\label{PGamma} P=\alpha \left[ \left( \frac{T_{e}}{T_{c^{\prime
}}}\right) ^{5}-\left( \frac{T_{ph}}{T_{c^{\prime }}}\right)
^{5}\right], \ee where $\alpha =4P_{0}~\Gamma (5)~\zeta (5)$,
$\zeta (x)$
 denotes Riemann's Zeta function and $\Gamma (n)$ is the Gamma function. A
comparison of Equation~(\ref{PGamma}) with the Equation (19) of Ref.~\cite{PBAllen}
shows that $P\propto T_{e}^{5}-T_{ph}^{5}$ is compatible with the $%
PDE$ in $SIN$ tunnel junctions [see
Refs.~\cite{FCWellstood,Pekola}]. This result shows that the usual
approach~\cite{caspek,caspek1,Martinis,Pekola} takes into account
only the net phonon flux. In this  approach details like
absorption at low frequency region and emission at high frequency
region~(see Fig.~\ref{SpectralAbsorption} ) are inaccessible.
However, the details of phonon absorption spectra are important
and may open new possibilities for cooling by $SIN$ tunnel
junctions as discussed in the next Section.

\section{Microrefrigerator based on the phonon deficit effect \label%
{PhononDeficitMicroRef}}

Why is it important to know the spectral dependence of the phonon
fluxes? How can the phonon-deficit-effect influence the
microrefrigerator design? The answers to these questions can be
found in \textit{spectral-selective} phonon filters.

\begin{figure}[h]
\par
\begin{center}
\leavevmode
\includegraphics[width=0.7\linewidth,height=0.5\linewidth]{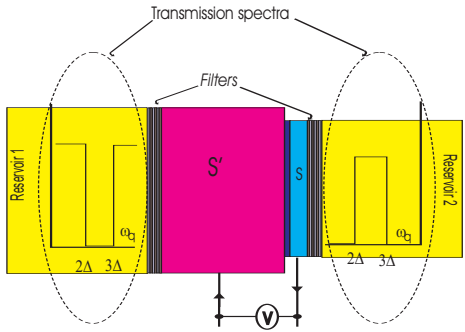}
\end{center}
\caption{Layout of the microrefrigerator with the improved properties.
Inserts show suggested spectral characteristics of superlattice filters. }
\label{fig_refr_design}
\end{figure}

To clarify this point, we should calculate the energy, which the
nonequilibrium phonons yield to the heatbath. This quantity is
equal to the integral of the spectral function plotted in
Fig.[\ref{figsub_gul_absorp}]. This integral having a net negative
value is likely to be higher in the case of an asymmetric
$S^{\prime }IS$ junction~\cite {GulianSergoyan,GulianMelkonyan}
and also in asymmetric $S^{\prime }IN$ junctions.
Fig.[\ref{figsub_gul_absorp}] displays the nonequilibrium phonon
flux from the $S^{\prime }IS$ tunnel junction, $\Delta
_{S}<<\Delta
_{S^{\prime }}$. \bigskip As follows from Figs.[\ref{figsub_gul_absorp}],~[%
\ref{AbsorbedT5}] and~[\ref{SpectralAbsorption}] the net flux integrated
over energy is negative~\cite{GulianSergoyan,GulianSergoyan2,GulianMelkonyan}%
. To enhance the $PDE$ effect and achieve better cooling, one should
separate nonequilibrium negative and positive fluxes across the interface
between the superconductor and the attached heat reservoir. Let us consider
placing a phonon spectral filter between superconductor $S$ and reservoir 2
(Fig.[\ref{fig_refr_design}]).
Suppose we can construct a filter~\cite {AGulian,GulianMelkonyan1}
which is transparent to phonons in a window of frequencies where
the flux in Fig.[\ref{figsub_gul_absorp}]
is negative. Such a filter would permit the flow of phonons ($\hbar \omega _{%
\mathbf{q}}\approx 2\Delta $) from the reservoir 2 to the superconductor $S$%
. At the same time the filter will absorb phonons below and above that
frequency window, including those coming from the heated barrier. Another
filter with complementary (i.e. the band pass) transmission properties may
be placed between reservoir 1 and the electrode $S^{\prime }$. It will
prevent phonons coming from $S^{\prime }$ (i.e., the reservoir 1) from being
absorbed in $S$. This will make the absorption from the reservoir 2 more
efficient. Then the system shown in Fig.[\ref{figsub_gul_absorp}] will work
as a refrigerator, by coolingthe reservoir 2. This cooling effect cannot
reduce the temperature much below the critical temperature of superconductor
$S^{\prime }$ (Fig.[\ref{figsub_gul_absorp}]). However, there are different
classes of superconductors with critical temperatures covering a wide range
of temperatures from about $150K$ to very low temperatures. Then, cooling
cascades can be organized in such a way to generate substantial cooling. We
suggest that such designs of a $PDE$-based microrefrigerator could be
carefully considered for the purpose of practical use. In the next section
we will discuss some designs.

\subsection{Design of phonon filters}

From the analysis in the preceding subsection it follows that the design of
appropriate phonon filters is of primary importance. Using the analogy
between reflection of a plane electromagnetic wave at the interface between
two optical media with different refractive indexes $n_{1}$ and $n_{2}$ and
the acoustic wave reflecting at the boundary of two elastic media with
different acoustic impendance $Z_{1}$ and $Z_{2}$, Narayanamurti et al.~\cite
{Narayanamurti,Koblinger} have developed appropriate phonon filters for
phonon spectrometery.

\bigskip

\begin{figure}[bth]
\par
\begin{center}
\leavevmode
\hspace{-1cm} \includegraphics[width=0.6\linewidth,height=0.6\linewidth%
,angle=-90 ]{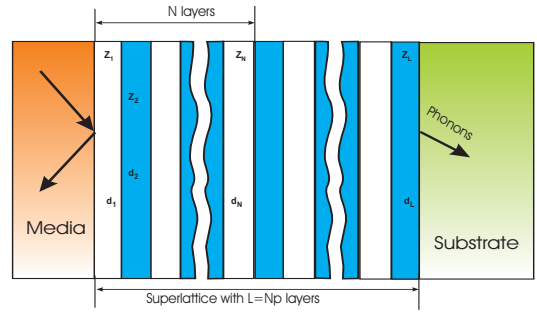}
\end{center}
\caption{Bragg multilayer between two media. }
\label{multiple_layer}
\end{figure}

\begin{figure}[bth]
\par
\begin{center}
\leavevmode
\includegraphics[width=0.48\linewidth,height=0.64\linewidth]{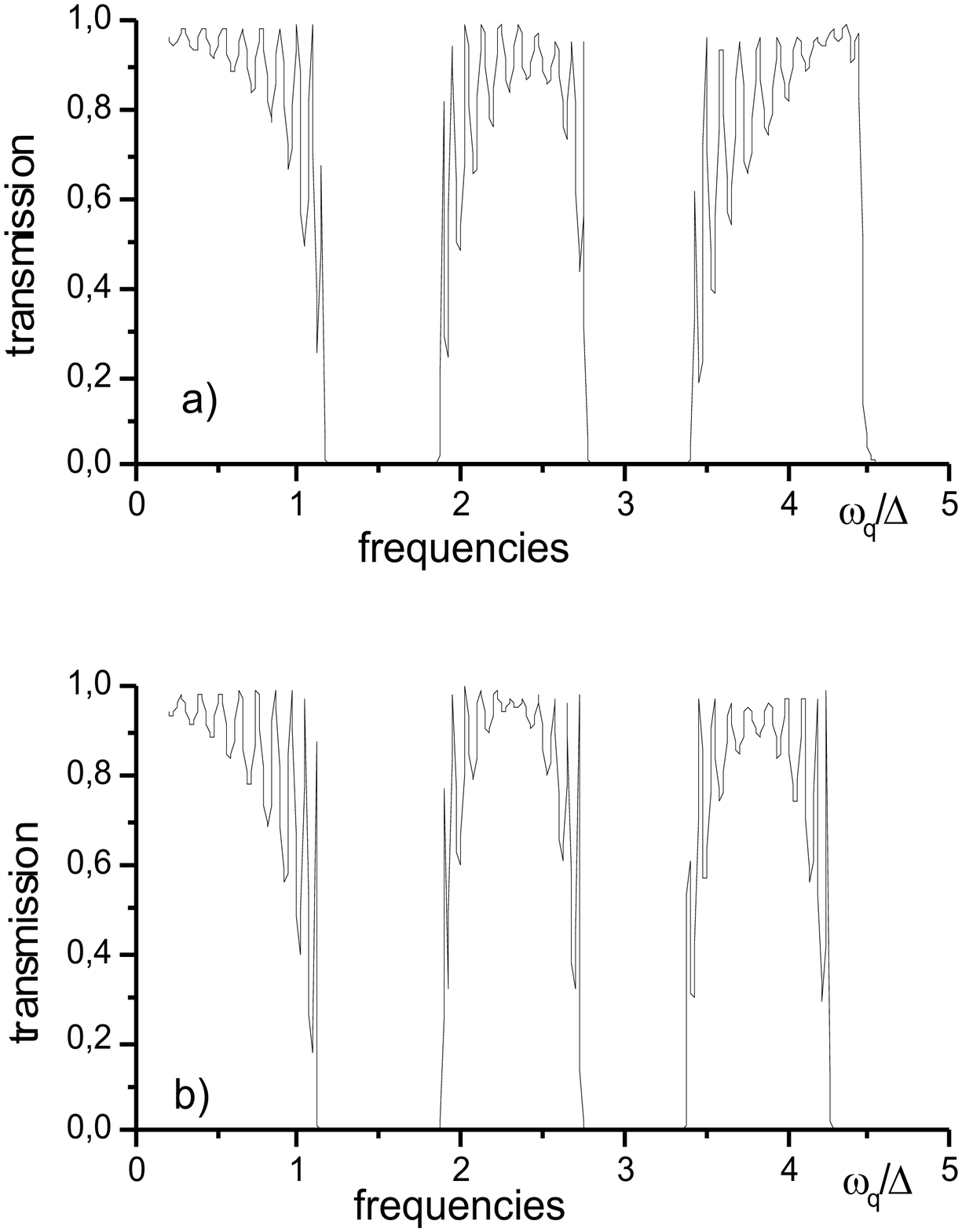} %
\includegraphics[width=0.48\linewidth,height=0.64\linewidth]{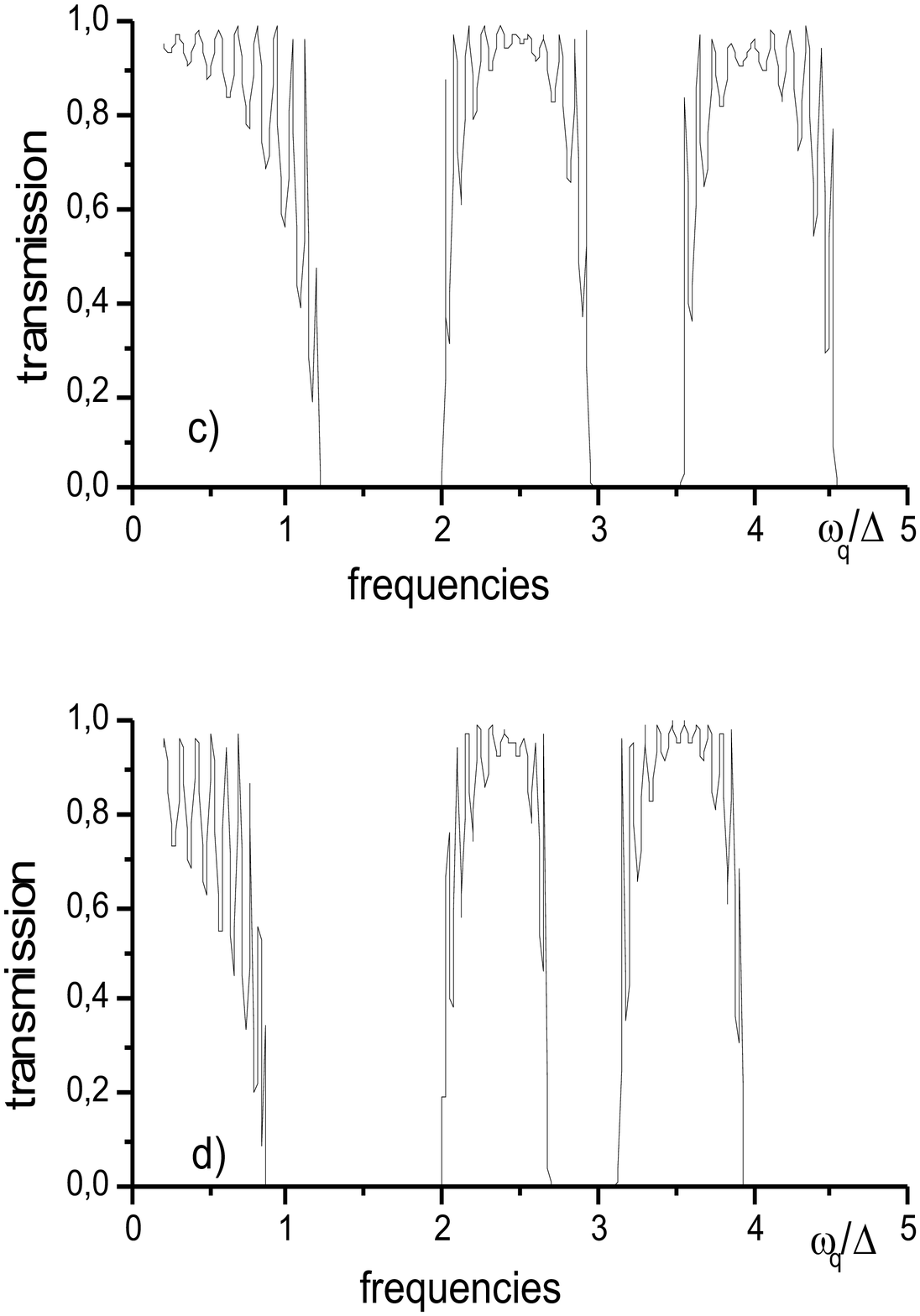}
\end{center}
\caption{Band pass properties of designed filters (for the parameters see
Tab.[\ref{Impendence}]). }
\label{pass_filter}
\end{figure}

Here we consider the simplest case of a phonon filter~\cite
{GulianMelkonyan,GulianMelkonyan1} for the case when acoustic
phonons propagate through a superlattice perpendicular to
interfaces. In Fig.[\ref{multiple_layer}] the superlattice system
is shown.
The system consists of a sequence of different layers: material 1 with
thickness $d_1$ and impedance $Z_1$, material 2 with thickness $d_2$ and
impedance $Z_2$, etc. If one repeats this configuration $p$ times then the
superlattice will have a period $D=d_1+d_2+\ldots+d_n$ and $L=Np$ layers.
The interfaces are perpendicular to the wave vector of the incident wave.
The superlattice is placed between two media with the acoustic impedance $%
Z_{medium}$ and $Z_{substrate}$. Methods of calculation for the transmission
and reflection coefficients in a linear stratified medium are described in
Refs.~\cite{Pierce,Richards,Shelkunoff,ShunLien}. We have presented results
of our calculation in Fig.[\ref{pass_filter}].
One should note that all filters have an oscillatory behavior for the
transmission coefficient in the band pass region. This is typical for band
pass filters. In Tab.[\ref{Impendence}] we list impedance and thickness of
the superlattices used in our calculations. Fig.[\ref{pass_filter}] presents
curves for the transmission spectra of the filter between reservoir 2 and
superconductor $S$ with different acoustic impedance, layer thickness and
layer number. Comparing with Fig.[\ref{figsub_gul_absorp}] we observe large
phonon fluxes at frequencies about $\omega_{\bq}=\Delta/\hbar$ and $\omega_{%
\bq}=3\Delta/\hbar$. The filter shown in Fig.[\ref{pass_filter}d] may work
well for those frequencies. The filter between reservoir 1 and
superconductor $S^\prime$ is presented in Ref~\cite{GulianMelkonyan1}. The
phonon filters for the $SIN$ tunnel junctions can be designed by a similar
analysis.
\begin{table}[bht]
\caption{ Parameters of the presented filters (see Figure~\ref{pass_filter}
). The impedance ($Z$) and the size of the layers ($d$) are given in units
of $10^5 g/(s ~ cm^2)$ and in Angstroms (\AA), respectively. }
\label{Impendence}
\begin{center}
\begin{tabular}{||c|c|c|c|c|c|c|c|c|c|c|c|c||}
\hline
& $Z_1$ & $Z_2$ & $Z_3$ & $Z_4$ & $Z_5$ & $d_1$ & $d_2$ & $d_3$ & $d_4$ & $%
d_5$ & N & p \\ \hline
a & 2.5 & 1.8 & 1.4 & 0.9 & 1.8 & 50 & 50 & 50 & 50 & 30 & 5 & 11 \\ \hline
b & 3 & 1.8 & 1.3 & 0.9 & 1 & 50 & 50 & 50 & 50 & 30 & 5 & 11 \\ \hline
c & 3 & 1.8 & 1.3 & 0.9 & 1 & 40 & 45 & 50 & 50 & 30 & 5 & 11 \\ \hline
d & 3 & 1.8 & 1.3 & 0.9 & 1 & 40 & 40 & 40 & 45 & 30 & 5 & 11 \\ \hline
\end{tabular}
\end{center}
\end{table}

\subsection{Phonon filters and refrigeration}

The proposed $PDE$ refrigerator is suited to refrigerate an object
attached to it. The effect of phonon filters can be estimated by
the value of $P/P_{0} $ from an experiment carried out by Pekola
et al~\cite{Pekola} where a membrane attached to the $SIN$ tunnel
junction was cooled. Our estimate for the net phonon absorption
$P/P_0$ by Equation~(\ref{PGamma}) varies from $0$ to
$10^{-2}$ depending on temperatures $T_e$ and $T_{ph}$. Table~\ref{Bgamma} presents our calculated values for
the nonequilibrium tunnel junctions (see also
Figure~\ref{SpectralAbsorption}). We find qualitative agreement
between experiment and calculation for the value of absorbed
power.
\begin{table}[tbh]
\caption{Maximal absorbed power $P/P_0 $ in the $SIN$ tunnel
junction at three different temperatures $T=0.27T_{c^{\prime
}},T=0.5T_{c^{\prime }}$ and $T=0.8T_{c^{\prime }}$.}
\label{Bgamma}
\begin{center}
\begin{tabular}{||c|c|c||c|c||c|c||}
\hline
$T/T_{c^\prime}$ & \multicolumn{2}{|c||}{0.27} & \multicolumn{2}{|c||}{0.5}
& \multicolumn{2}{c||}{0.8} \\ \hline
$V/T_{c^\prime}$ & \multicolumn{2}{|c||}{$1.56$} & \multicolumn{2}{|c||}{$%
1.31$} & \multicolumn{2}{|c||}{0.79} \\ \hline
$\nu/\gamma$ & 0.005 & 0.02 & 0.005 & 0.02 & 0.005 & 0.02 \\ \hline
$P/P_0$ & $0.004$ & $0.014$ & $0.0052$ & $0.0224$ & $%
0.0011$ & $0.0042$ \\ \hline
\end{tabular}
\end{center}
\end{table}

\begin{figure}[bth]
\par
\begin{center}
\leavevmode
\includegraphics[width=0.82\linewidth,height=0.92\linewidth]{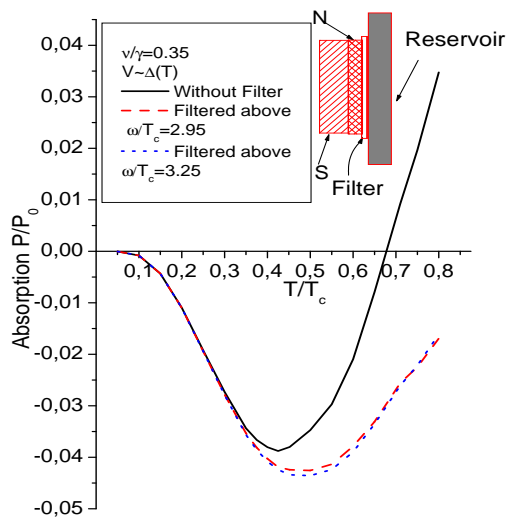} %
\end{center}
\caption{ Phonon flux with low pass filter at $\omega=2.95T_c$ and
$\omega=3.25T_c$ } \label{pass_filtered}
\end{figure}

The phonon filters will increase the absorbed power by cutting off
the emission window (see Figure~\ref{SpectralAbsorption}).
Table~\ref {AfterFilter} presents the absorbed power after placing
a low pass filter (see Figures~\ref{SpectralAbsorption}
and~\ref{pass_filtered}) between the $SIN$ tunnel junction and an
object. Phonon absorption is increased at $T/T_c=0.5$. At
$T/T_c=0.8$ we see that $SIN$ tunnel junction with phonon filter
will absorb phonons (the minus sign in Table~\ref{AfterFilter}
means absorption).
\begin{table}[tbp]
\caption{Absorbed power $P/P_0$ in the $SIN$ tunnel junction at
two different temperatures $T=0.5T_{c^\prime}$ and
$T=0.8T_{c^\prime}$, without and with phonon filter. The data
below correspond to Fig.~[\ref {SpectralAbsorption}]. The ratio
$\nu/\gamma=0.02$.} \label{AfterFilter}
\begin{center}
\begin{tabular}{||c|c|c||c|c||}
\hline
$T/T_{c^\prime}$ & \multicolumn{2}{|c||}{0.5} & \multicolumn{2}{c||}{0.8} \\
\hline
$V/T_{c^\prime}$ & \multicolumn{2}{|c||}{$1.31$} & \multicolumn{2}{|c||}{$%
1.31$} \\ \hline
effect & total & filtered & total & filtered \\ \hline
$P/P_0$ & $-2.24\,10^{-2}$ & $-2.43\,10^{-2}$ & $1.24\,10^{-4}$ & $%
-2.5\,10^{-2}$ \\ \hline
\end{tabular}
\end{center}
\end{table}
Fig.~[\ref{pass_filter}] shows the temperature dependence of the
total phonon absorption $P/P_0$ for three different casses. The
solid line is the phonon absorption by a $SIN$ tunnel junction
without phonon filter. The dotted and dashed curves present the
filtered phonon absorption for two different low pass filters with
transmittance $1$ at $\omega<\,2.95T_c$ and $\omega< \,3.25T_c$.
Thus the phonon absorption is incresed in a wide temperature
region.

We recall that the refrigeration schema presented above is suited
to refrigerate an object attached to it. But as follows from the
phonon absorption spectra an analogous design with high pass
filters can prevent the phonon absorption by $SIN$ tunnel junction
and thus reduce the heating of electrons by phonons (see
Figures~[\ref{SpectralAbsorption},~\ref{pass_filtered}]).

\section{Conclusions \label{Concl}}

The $SSMR$ performance depends on various external and internal parameters.
The $PDE$-based approach presented here and the usual apprach~\cite
{PBAllen,FCWellstood,Pekola} give the similar expression (Eq.~\ref{PGamma})
for the net cooling in case of $SIN$ tunnel junction. Thus one can conclude
that the cooling of a membrane in the experiment by Pekola et al.~\cite
{Pekola} is a direct implication of the phonon deficit effect. The $PDE-$%
approach offers explicit spectral description for participating phonon
fluxes. It provides receipes to overcome unwanted heating channels and shows
new details which have been omitted in a simplified ''cooling by
extraction'' description.

We have analyzed the ``phonon deficit effect'' in $SIN$ tunnel junctions. In
contrast to $SIS$ tunnel junctions there is a temperature region where the $%
SIN$ tunnel junctions essentially absorb low frequency phonons. As
temperature is increased, the $SIN$ tunnel junctions also emit phonons to
the environment.
Depending on the value of the ratio $\nu /\gamma $ there is net phonon
emission or absorption. The temperature dependence of the net phonon
absorption is consistent with experimental observations~\cite
{Pekola,caspek,Martinis}: there is heating at high and low temperatures and
refrigeration at intermediate temperatures.

To enhance the cooling power of the $SIS$($N$) and tunnel
junctions and overcome the phonon heating at high temperatures,
phonon filters can be effective. We performed calculations for the
microrefrigerator phonon filters. Those filters have superlattice
design and can be obtained using
contemporary nanotechnology methods. They can enhance the cooling power of $%
SIS$ and $SIN$ tunnel junctions.

\vspace{0.5cm}

\begin{acknowledgments}
\noindent H.K. has been supported by NSERC Canada.
G.M. is grateful for support by
the ``Programme qu\'eb\'ecois de bourses d'excellance'' of the governement of Qu\'ebec.
\end{acknowledgments}

\appendix*
\section{Phonon flux between electrons and Phonons \label{derivation}}

In this appendix, we give the derivation of the
expression~(\ref{PGamma}).

 The general expression for the phonon flux is given by (compare with the
 formula~(\ref{PhononEmmison}), $\hbar=1$,$k_B=1$)
\begin{equation}
\label{PhononEmmisionAppendix} {P}=\int_0^{\omega_D} \omega_q ~
\rho_0(\omega_q) ~
I^{ph-el}(N_{\omega_q}^0){{\mathrm{d}}\omega_q}.
\end{equation}
Here $I^{ph-el}(N_{\omega_q}^0)$ is the phonon-electron coillision integral
and it is given by
\begin{eqnarray}
\label{PhononElectronCollision}
I^{ph-el}\left( N_{\omega_q}\right)&=&\frac{\pi \lambda \omega_D}{8
\epsilon_F}\int_{0}^{\infty} \mathrm{d}\epsilon \int_{0}^{\infty}
\mathrm{d} \epsilon^\prime \left\{ {\cal
S}_1\delta\left(\epsilon+\epsilon^\prime -\omega_q \right) \right. \nonumber\\
&& +\left. 2{\cal S}_2\delta
\left(\epsilon-\epsilon^\prime -\omega_q \right) \right\},
 \end{eqnarray}
where  ${\cal S}_1$ and ${\cal S}_2$ are elementary collison
processes. When the distribution function of electrons and holes
is Fermi function then $n_{\epsilon}=n_{-\epsilon}$. In this case
${\cal S}_1$ and ${\cal S}_2$ can be written as
\begin{equation}
\label{S1} {\cal S}_1(\epsilon,\epsilon^\prime,\omega_q)=4\left[
\left( N_{\omega_q}+ 1\right) \npe\npepr -
N_{\omega_q}\left(1-\npe\right)\left(1-\npepr\right) \right],
\end{equation}
\begin{equation}
\label{S2} {\cal S}_2(\epsilon,\epsilon^\prime,\omega_q)= 4\left[
\left( N_{\omega_q}+1\right) \npe\left(1-\npepr\right) -
N_{\omega_q}\left(1-\npe\right)\npepr \right].
\end{equation}
 After
substitutions of the expressions~(\ref{S1}),~(\ref{S2}) and
(\ref{PhononElectronCollision}) in the
expression~(\ref{PhononEmmisionAppendix}), the integration by
delta function gives
\begin{eqnarray}
{P}&=&\frac{4\pi\lambda\omega_D}{8\epsilon_F}\int_0^{\omega_D}{{\mathrm{d}}\omega_q}
\omega_q ~ \rho_0(\omega_q)\left[ \int_0^{\omega_q}~
{\cal S}_1(\epsilon,\omega_q-\epsilon,\omega_q){~\mathrm{d}\epsilon}\right.\nonumber\\
&&+\left.\int_{\omega_q}^{\omega_D+\omega_q}~ 2{\cal
S}_2(\epsilon,\epsilon-\omega_q,\omega_q){~\mathrm{d}\epsilon}\right].
\end{eqnarray}
Finally evaluating these integrals  by noting that $T\ll\omega_D$
and using the integral representation of Rieman zeta
function~$\zeta(n+1)\Gamma(n+1)=\int_0^{\infty}\,x^n/(exp(x)-1)\,d\,x$
one obtains \be {P}=4\frac{\Omega}{16\pi}\frac{\lambda
\omega_D}{\epsilon_F u^3}\Gamma
(5)\zeta(5)\left(T_e^5-T_{ph}^5\right). \ee

\bibliographystyle{plain}

\end{document}